# On Integrating Design Thinking for a Human-centered Requirements Engineering

Jennifer Hehn, Daniel Mendez, Falk Uebernickel, Walter Brenner, Manfred Broy

**Abstract**— In this position paper, we elaborate on the possibilities and needs to integrate Design Thinking into Requirements Engineering. We draw from our research and project experiences to compare what is understood as Design Thinking and Requirements Engineering considering their involved artifacts. We suggest three approaches for tailoring and integrating Design Thinking and Requirements Engineering with complementary synergies and point at open challenges for research and practice.

◆ ———————————

## 1 INTRODUCTION

REQUIREMENTS ENGINEERS often face the challenge of discovering and satisfying the fuzzy needs and volatile requirements of the various stakeholders involved. Design Thinking, as a human-centered, rapid-prototyping method for innovative design, is one promising approach to address this challenge [1], [2].

We postulate that we need an effective integration of Design Thinking and Requirements Engineering. However, little is known how an integration could be realized considering a holistic view, also due to existing misconceptions: In Design Thinking, we pretend too often that problem-solving ends with understanding the problem and building a non-technical prototype, leaving open the seamless transition into software development endeavors. In Requirements Engineering, we pretend too often that software requirements are somehow "just there" and simply need to be elicited, missing the potential of fully exploring the problem space. One difficulty to be taken into account is that, like with other "agile" approaches, Design Thinking can appear as a set of single methods, tools, or even as a holistic approach [3]. Requirements Engineering, in turn, is an engineering discipline encompassing various principles, tools, and even more methods – all to be selected depending on given project situations and software process models [4]. To make effective use of the full potential of Design Thinking, we first need a better understanding of what it is and how it relates to Requirements Engineering, in which situations it might be suitable, and how it could be properly integrated.

For years, we accompanied and researched organizations that adopted Design Thinking in their industrial settings. In this position paper, we share our experiences and outline synergies and differences between Design Thinking and Requirements Engineering with a model of artifacts that emerges from industrial adoptions. We recommend three integration strategies before concluding with open questions for research and practice.

## 2 DESIGN THINKING IN A NUTSHELL

Design Thinking is a structured problem-solving approach to develop innovative products, services, and business models. It builds upon the exploration of human needs, non-technical prototyping, iterative problem reframing, and interdisciplinary teamwork. Design Thinking is primarily intended to be applied in "wicked" project settings which are characterized by volatile and partially unidentified/hidden requirements [5]. Our own scope and experiences are centered in the development of software-intensive products in industrial projects as well as in practical courses at the University of St.Gallen [6]. There, we employ an iterative and multi-disciplinary approach widely known as the Design Thinking "double diamond" shown in Fig. 1.

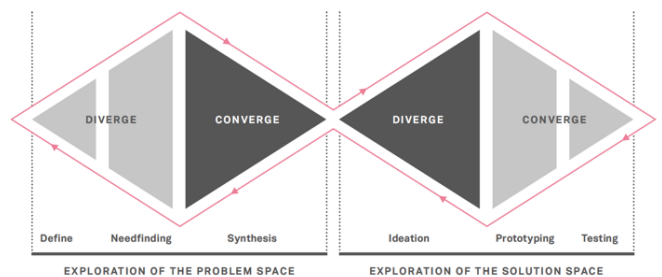

Fig. 1. Design Thinking Double Diamond

---

- *Jennifer Hehn is with the Institute of Information Management, University of St.Gallen, Switzerland. E-mail: jennifer.hehn@unisg.ch.*
- *Daniel Mendez is with the Blekinge Institute of Technology and with fortiss GmbH, E-mail: daniel.mendez@bth.se.*
- *Falk Uebernickel is with the Hasso-Plattner Institute at the University of Potsdam, Germany and the University of St.Gallen, Switzerland. E-mail: falk.uebernickel@hpi.de.*
- *Walter Brenner is with the Institute of Information Management, University of St.Gallen, Switzerland. E-mail: walter.brenner@unisg.ch.*
- *Manfred Broy is with the Technical University of Munich (TUM). E-mail: broy@in.tum.de.*

We distinguish the problem space from the solution space, each with exploratory (diverging) and defining (converging) activities. The problem space contains methods to capture the problem in a human-centric, empathic manner. The phases *Define* and *Needfinding* explore the user and business environment, while *Synthesis* condenses the gathered information to potential opportunities for the discovered needs. The solution space contains methods to develop *ideas*, build *prototypes*, and systematically *test* them. Iterations are carried out wherever necessary in the process. Prototypes evolve from rudimentary, and often paper-based, low-fidelity prototypes to more sophisticated, technical ones at later stages. The ability to conduct this stepwise improvement of assumptions, ideas, and prototypes fundamentally relies on an open communication environment which is leveraged by harnessing selected tools and techniques like the ones summarized in our online material compendium (www.dt4re.org).

## 3 CROSS COMPARISON OF DESIGN THINKING AND REQUIREMENTS ENGINEERING

### 3.1 Two Complementary Approaches

To compare both approaches, it needs to be clear what is being achieved through Design Thinking and what through Requirements Engineering and how the output can be transformed into an actual product. Based on our experiences as practitioners who train companies in Design Thinking and Requirements Engineering [7], [8], we created a blueprint of relevant artifacts from both approaches (Fig. 2). The artifacts describe the produced work results and their dependencies and, thus, abstract from complex development processes, which are barely comparable across projects [9].

Our model structures the artifacts according to why a system is needed (*context layer*), which (user-level) requirements and features are necessary (*requirements layer*), and how the system is to be realized (*system layer*). The full model consists of 40 artifacts: 16 attributed to Design Thinking, 16 to Requirements Engineering, and 8 to both (for details, see *Web Extra*).

We see various commonalities between Design Thinking and Requirements Engineering, at least if the latter is understood as an iterative approach. The differences should be seen as complementary activities. Design Thinking expands the toolbox for Requirements Engineering by emphasizing artifacts describing the relevance of the system vision (*context layer*). They complement the more technical-oriented Requirements Engineering artifacts with a human-centered perspective [10]. And, Requirements Engineering expands the toolbox of Design Thinking. The technical realization of the functionalities is not within the scope of Design Thinking and is rather attributed to the artifacts produced in Requirements Engineering (*requirements* and *system layer*). Based on our model we suggest the following:

**Use Design Thinking to guide your requirements elicitation.** The Design Thinking process model (Fig. 1) can help to produce relevant *context* artifacts for a comprehensive understanding of the problem. Use the high-fidelity prototype to visualize the system vision including key functionalities and the general form of user interaction. In addition, the prototype serves usability-driven demonstration purposes to customers rather than technical-driven feasibility studies.

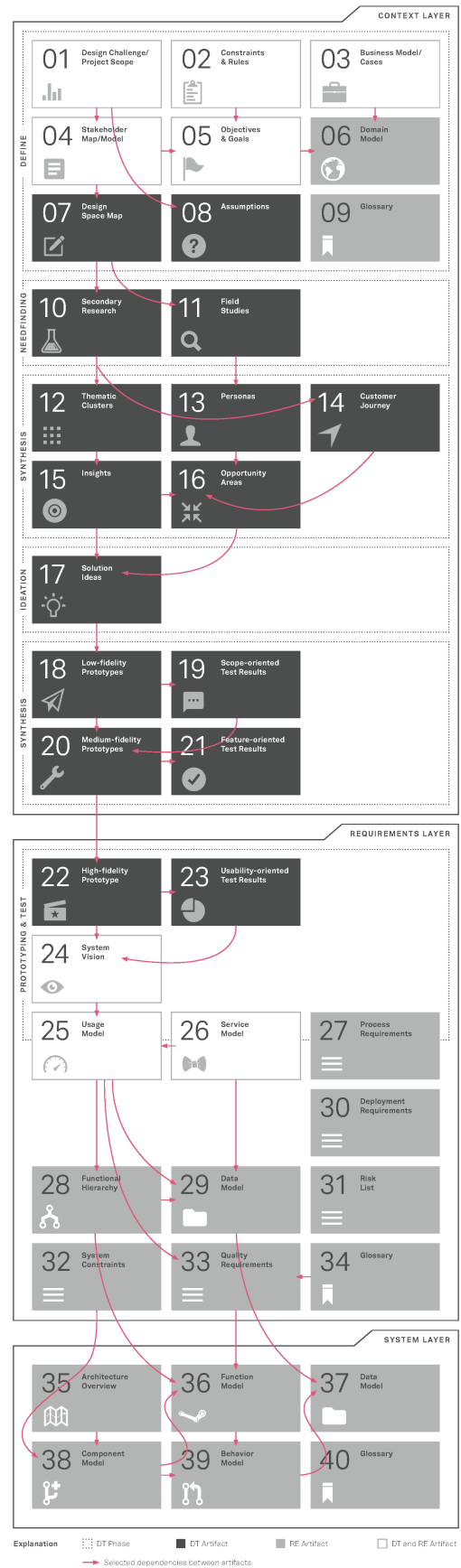

Fig. 2. Combined Artifact Model of Design Thinking and Requirements Engineering

**Embrace the learning curve of Design Thinking to inform implementation.** From an implementation point of view, understanding how ideas and functionalities have emerged, helps to design the solution vision in the intended form. For example, field study results and insights can inform use cases and scenarios – typically defined in Requirements Engineering (or user stories when working in a more agile way).

**Define your content-focus.** Depending on what a team needs to learn more about, each approach emphasizes a different content. If you need to understand the user and business context in detail, concentrate on Design Thinking artifacts; if you focus on the technical perspective and feasibility questions, concentrate on the Requirements Engineering artifacts at the requirements and system layers. Teams may jump back and forth between both approaches, if new questions come up in one or the other area.

**Find the right balance.** A balance needs to be found between creative and corporate requirements. Experienced practitioners are able to observe different demands and adapt the approach accordingly. The differences in Design Thinking and Requirements Engineering are therefore a great chance to customize projects along the way.

### 3.2 Three Integration Strategies

We have identified three valuable strategies to customize projects based on different ways to integrate Design Thinking into Requirements Engineering: (1) run Desin Thinking prior to performing Requirements Engineering practices (*upfront*), (2) infuse the existing Requirements Engineering process with selected Design Thinking tools and artifacts (*infused*), or (3) integrate Design Thinking into Requirements Engineering practices on an ongoing basis (*continuous*). Tab. 1 provides an overview when and how each strategy can be operationalized best. To choose the right strategy consider your objective and project context.

*Use the upfront strategy,* when you are facing a high level of uncertainty about the problem and the solution. Applying Design Thinking helps to understand the problem deeply and define the overall concept of an idea.

*Use targeted infusions* to support existing Requirements Engineering activities, for example, if you already have a solution idea, but still need to sharpen it or if you want to clarify fuzzy requirements. Applying Requirements Engineering helps to focus on details or features of an idea.

*Use continuous Design Thinking,* when integrating Design Thinking principles into your Requirements Engineering routine, for example as part of an organizational change program. Instantiate roles (human-centric requirements engineers) and use the Design Thinking principles as a longterm strategy.

We have also found the maturity level of Design Thinking in an organization as an influencing factor. While Requirements Engineering is typically a common practice, Design Thinking is still relatively new. Thus, the decision to integrate both approaches also depends on the level of courage, given time, and dedicated resources. As a rough guideline, the infusion strategy provides a reasonable starting point as it applies focused Design Thinking interventions within established practices. While the upfront strategy also keeps existing procedures, it requires more time and resources. Finally, the continuous strategy demands for management commitment to foster mindset change in an organization or department.

## 4 OPEN CHALLENGES

We summarized three strategies to effectively connect Design Thinking and Requirements Engineering. Applying Design Thinking in an upfront manner or a way in which it co-exists with engineering activities is what we typically encounter in practice. A fully integrated and continuous Design Thinking is, however, what we need to facilitate seamless transitions into engineering activities (and back). We showed one such integration at an artifact level which raises us to the next level of challenges for research and practice. This is also because we cannot yet unfold a complete picture of how principles, work results, and methods as found in Design Thinking and other Software Engineering practices (beyond Requirements Engineering) exactly relate to each other.

At a **conceptual level**, we still need to better understand two major aspects: 1) Which principles in Design Thinking can also be found in other more holistic human-centered software engineering disciplines and how do these differ? 2) What are their boundary objects? This becomes apparent in our artifact model: What are the same or similar artifacts with which purposes? When are they interchangeable? Same holds when reflecting upon the methods used to create the artifacts: Which methods in Design Thinking can be used for other software engineering disciplines? How do these methods differ and how can they be integrated? How can milestones be effectively defined, for instance, as interfaces between different software process models? Finally, we also need to reflect upon project roles. What seems trivial at first becomes challenging when considering competencies and responsibilities. How can, for instance, multidisciplinary Design Thinking teams be integrated with traditional roles such as the one of a requirements engineer or a business analyst? How do existing responsibilities have to be modified when co-existing and collaborating?

Further questions arise when putting an integration into action **at the project level**: What are typical project situations which influence the choice of a strategy? How do these situations and the class of systems influence the choice of a strategy and/or single methods? How can these situations be characterized and assessed (with which confidence)? The latter is essential to eventually build a holistic approach tailorable to the needs of individual software project settings and, thus, ready for adoption in industry.

TABLE 1
STRATEGIES FOR INTEGRATING DESIGN THINKING (DT) AND REQUIREMENTS ENGINEERING (RE)

| | UPFRONT DESIGN THINKING | INFUSED DESIGN THINKING | CONTINUOUS DESIGN THINKING |
|---|---|---|---|
| Characteristics | - Used at an early project stage to provide clarity for unclear user needs and to define a solution vision<br>- Uses DT as guiding process<br>- Requires setup of a (mini-)project | - Used within an existing RE process to gain new ideas or clarify fuzzy requirements<br>- Uses DT as toolbox in RE<br>- Requires setup up of focused workshops | - Continuous usage of DT and RE elements to realize end-2-end view from user need to solution vision to functional solution<br>- Uses DT as guiding mindset<br>- Combines upfront and infused strategy and setup of new role |
| Outcome | - Clear solution vision in form of a mockup<br>- Comprehensive set of DT artifacts (Fig. 2) as basis to perform further RE activities | - Outcome is situation-dependent (e.g. (new) features, user requirements, test with users – all along RE process)<br>- Selected set of DT artifacts (Fig. 2) | - Software Requirements Specification based on and tracebale to customer needs<br>- Comprehensive set of DT and RE artifacts as shown in Fig.2 |
| Duration | - Approximately 3-12 weeks | - Approximately ca. 1-10 days | - More than 6 months |
| Key Roles | - DT Team (4-6) from different areas of expertise<br>- Extended Team of experts and project sponsor: offers internal expertise and defines initial design challenge<br>- DT Coach: provides process guidance | - Workshops participants (5-20) from different areas of expertise<br>- DT Coach: provides process guidance<br>- Requirements Engineer as mediator between DT Team and Software Development Team | - See upfront and infused strategies<br>- Human-centric Requirements Engineer: New role incorporating DT expertise as well as RE expertise and mediating between both schools of thoughts. |
| Benefits | - Full potential of DT is leveraged while no changes to RE are necessary<br>- Solution concept has traceable links to user customer needs<br>- Deep context understanding is achieved | - Only minimal changes in existing RE practices are required<br>- Resource- and time-friendly<br>- Low adoption hurdle for DT | - Seamless integration incl. development-critical artifacts<br>- Human-centred mindset throughout entire project<br>- Precise and traceable (user) requirements through continuous identification of new requirements and testing |
| Challenges | - Resource- and time-intense<br>- Lost implicit knowledge when handing-over prototype<br>- No to little attention on development-critical artifacts | - Risk of neglecting problem context<br>- Missing sustainable impact of DT<br>- No to little attention on development-critical artifacts | - Commitment-, resource- and time-intense<br>- Highly team-dependent<br>- Requires organizational mind-shift and support |
| Case Example | The international Alpha Insurance company wanted to develop a new service for their new target group of "young professionals". A project team from five different business functions (Marketing, IT, Actuary, Product Manager, Claims) with 40% capacity followed the DT process in an iterative manner for three months. The solution vision resulted in a tested medium-fidelity prototype for a digital on-demand insurance that could be activated and deactivated based on the user's preferences. The DT team handed over the prototype to the implementation team for further specification, testing, development, and market introduction. Transferred artifacts included a project documentation with 20 field studies, 2 personas, 5 opportunity areas, and 6 low-fidelity prototypes with learnings about failures. The final solution vision (in form of a mockup) specified key features and their usability. The implementation team performed tests to validate these features, their usability, and their service model. | Beta Enterprises is an international electronics group that wanted to evaluate the possibilities of smartphone applications for container ships in a marine context. The main goal was to define requirements from a user point of view and to foster creativity for solution finding. In a highly regulated environment, a DT infusion was chosen to support the ongoing RE activities with selected tools from needfinding and prototyping. Five DT infusion sessions (1-2 days) were conducted within five months. Produced artifacts included field studies for precise user requirements (it was the first time the team had gotten in close contact with marine captains) and tested medium-fidelity prototypes to sharpen service and usage models. According to the workshop participants having direct user contact raised their confidence level in the success of the intended solution. Initial concerns about not being able to find interview partners in a highly sensitive B2B setting turned out as unjustified. | Gamma Energy is a large energy provider with subsidiaries worldwide. A diverse project team applied an upfront DT approach to explore the potential of platforms in the utility sector. The outcome was a solution vision for a digital home improvement platform to advance lead generation. To ensure a human-centered mindset throughout specification and development, a new role was instantiated to use selected DT tools to enhance the prototype and fill the backlog with new features. Produced DT artifacts included high-fidelity prototypes with usability- and feature-oriented test feedback and new solution ideas. For development Scrum became the guiding framework that enabled the entire project team to work in sprints. During development DT prototypes were used as boundary objects to enhance communication with relevant internal stakeholders and to foster a human-centred mindset within the team. |

## 5 CONCLUSION

We have drawn from our experiences to discuss how Design Thinking can be used effectively for Requirements Engineering. Both approaches aim at discovering goals and requirements. While both, Design Thinking and Requirements Engineering, are very distinct when it comes to the underlying philosophies, many artifacts are complementary or even overlapping. Yet, while in Requirements Engineering, the measurements of success are often the documented requirements as a foundation for development and quality assurance, in Design Thinking, we follow a philosophy of domain understanding and the learning curve leading to it – regardless the surrounding processes. We showed different combination strategies depending on the project context before we laid out a roadmap for future research and practice. With this manuscript, we hope to foster an important and overdue discussion and we cordially invite researchers and practitioners to join this endeavour with their own ideas of integrating Design Thinking for a human-centered approach to Requirements Engineering.

**Jennifer Hehn** is a research associate, pursuing her PhD in business innovation, at the University of St.Gallen's Institute of Information Management. Her research interests are Design Thinking, Requirements Engineering, and agile development techniques. She is also a senior manager at the innovation consultancy IT Management Partner St.Gallen AG (ITMP) with the focus on managing Design Thinking projects in a variety of industries.

**Daniel Mendez** is associate professor for software engineering at the Blekinge Institute of Technology, Sweden, and senior scientist at fortiss, the research institute of the Free State of Bavaria for software-intensive systems and services. His research is on Empirical Software Engineering with a particular focus on interdisciplinary, qualitative research in Requirements Engineering and its quality improvement. Further information is available at http://www.mendezfe.org

**Falk Uebernickel** is professor at the Hasso-Plattner-Institute in Potsdam and adjunct professor at the University of St.Gallen responsible for Design Thinking and Business Innovation. He has worked together with and consulted several international organizations in implementing Design Thinking.

**Walter Brenner** is professor at the University of St.Gallen and acts as Managing Director of the Institute of Information Management and the Institute of Computer Science. His research focuses on Industrialisation of Information Management, Management of IT service providers and Innovation and Technology Management. Prior to joining academia Professor Brenner worked as Head of Application Development with Alusuisse-Lonza AG.

**Manfred Broy** is full professor emeritus for software & systems engineering at the Technical University of Munich. His research interests are software and systems engineering comprising both theoretical and applied aspects including system models, specification and refinement of system components, specification techniques, development methods and verification. Further information is available at https://www4.in.tum.de/~broy/